\documentclass[aip,apl,twocolumn,reprint]{revtex4-1}
\usepackage{graphicx}% Include figure files
\usepackage{dcolumn}% Align table columns on decimal point
\usepackage{bm}% bold math
\usepackage{color}
\usepackage{tabularx}
\usepackage{array}
\usepackage{amsmath}
\usepackage{pbox}
\usepackage{stmaryrd}  %\llbracket \rrbracket

\begin{document}

\title{Experimental parameters, combined dynamics, and nonlinearity of a Magnonic-Opto-Electronic Oscillator (MOEO)}

\author{Yuzan Xiong}
\affiliation{Department of Physics, Oakland University, Rochester, MI 48309, USA}
\affiliation{Department of Electronic and Computer Engineering, Oakland University, Rochester, MI 48309, USA}
\affiliation{Materials Science Division, Argonne National Laboratory, Argonne, IL 60439, USA}

\author{Zhizhi Zhang}
\affiliation{Materials Science Division, Argonne National Laboratory, Argonne, IL 60439, USA}

\author{Yi Li}
\affiliation{Materials Science Division, Argonne National Laboratory, Argonne, IL 60439, USA}

%\author{Rao Bidthanapally}
%\affiliation{Department of Physics, Oakland University, Rochester, MI 48309, USA}

\author{Mouhamad Hammami}
\affiliation{Department of Physics, Oakland University, Rochester, MI 48309, USA}

%\author{Ian McKinnon}
%\affiliation{Department of Physics, Oakland University, Rochester, MI 48309, USA}

\author{Joseph Sklenar}
\affiliation{Department of Physics and Astronomy, Wayne State University, Detroit, MI 48201, USA}

\author{Laith Alahmed}
\affiliation{Department of Electrical and Computer Engineering, Auburn University, Auburn, AL 36849, USA}

\author{Peng Li}
\thanks{The authors to whom correspondence may be addressed: peng.li@auburn.edu, and, }
\affiliation{Department of Electrical and Computer Engineering, Auburn University, Auburn, AL 36849, USA}

%\author{John Pearson}
%\affiliation{Materials Science Division, Argonne National Laboratory, Argonne, IL 60439, USA}

\author{Thomas Sebastian}
\affiliation{THATec Innovation GmbH, Augustaanlage 23, 68165 Mannheim, Germany}

%\author{Gopalan Srinivasan}
%\affiliation{Department of Physics, Oakland University, Rochester, MI 48309, USA}

\author{Hongwei Qu}
\affiliation{Department of Electronic and Computer Engineering, Oakland University, Rochester, MI 48309, USA}

\author{Axel Hoffmann}
\affiliation{Department of Materials Science and Engineering, University of Illinois at Urbana-Champaign, Urbana, IL 61801, USA}

%\author{Guangzhi Qu}
%\affiliation{Department of Computer Science and Engineering, Oakland University, Rochester, MI 48309, USA}

\author{Valentine Novosad}
\affiliation{Materials Science Division, Argonne National Laboratory, Argonne, IL 60439, USA}

\author{Wei Zhang}
\thanks{weizhang@oakland.edu}
\affiliation{Department of Physics, Oakland University, Rochester, MI 48309, USA}

\date{\today}

\begin{abstract}

We report the construction and characterization of a comprehensive magnonic-opto-electronic oscillator (MOEO) system based on 1550-nm photonics and yttirum iron garnet (YIG) magnonics. The system exhibits a rich and synergistic parameter space because of the ability to control individual photonic, electronic, and magnonic components. Taking advantage of the spin wave dispersion of YIG, the frequency self-generation as well as the related nonlinear processes become sensitive to the external magnetic field. Besides being known as a narrowband filter and a delay element, the YIG delayline possesses spin wave modes that can be controlled to mix with the optoelectronic modes to generate higher-order harmonic beating modes. With the high sensitivity and external tunability, the MOEO system may find usefulness in sensing applications in magnetism and spintronics beyond optoelectronics and photonics.

\end{abstract}

\maketitle

\section{Introduction}

Auto-oscillation is a topic studied in diverse branches of science including nonlinear dynamics \cite{markus_rmp2014,mikhail_rmp2006,hassan_rmp2017}, chaos \cite{argyris_nature2005}, neuroscience \cite{varela_natrevneuro2001}, and laser theory \cite{haken_ZPhys1963}. It is a topic which combines physics and mathematics with a great amount of applications. In particular, synchronization in an auto-oscillator network is believed to be a promising contender for future neuromorphic engineering \cite{grollier_jap2018}, where many demonstrations have been found in physics and microwave electronics such as Josephson junctions \cite{chernilov_pre1995} or spin-torque nano-oscillators (STNO) \cite{slavin_ieee2009,ruotolo_nnano2009,grollier_prb2006,vasyl_apl2009}. 

A steady-state auto-oscillation involves energy compensation by an active element (negative damping) to a dissipating element (positive damping), which, in a general case, are both nonlinear \cite{slavin_ieee2009}. In many cases the process of energy transfer, from the active to the passive elements, may take much more time than the oscillation period. Therefore, the dynamics of the auto-oscillation can be strongly influenced by any time delay existing in the system. In fact, many of the practically interesting auto-oscillating systems, such as an optoelectronic oscillator (OEO) \cite{yao_ieee1996,miguel_rmp2013,larger_rtsa2013} and a microwave feedback ring \cite{wang_prl2011,wu_ssp2010}, are both non-isochronous and rely on a delayed feedback. 

The principle of operation of OEOs is based on a circulation of microwave signals in a loop circuitry consisting of microwave and optical paths, which are typically integrated by using a Mach-Zehnder Modulator (MZM), i.e., an optical device capable of imposing modulation on the phase, frequency, amplitude, or polarization of the telecommunication wavelengths, taking advantage of the electro-optic effects of lithium niobate crystals. \cite{yao_ieee1996,miguel_rmp2013,larger_rtsa2013} In past decades, many advances have been made for optoelectronic microwave oscillators, driven by increasing demands and rapid advances in radio-frequency (rf) technologies, radar and telecommunication systems. \cite{lute_npho} From a scientific point of view, OEO systems with a delayed feedback loop have become a useful platform for studying nonlinearity, chaos, and synchronization of microwaves and optics. In addition, its mathematical extracts can be readily applied to a range of physical, electronic, and biological systems \cite{yanne_rmp2019}. 

\begin{figure*}[htb]
 \centering
 \includegraphics[width=6.5 in]{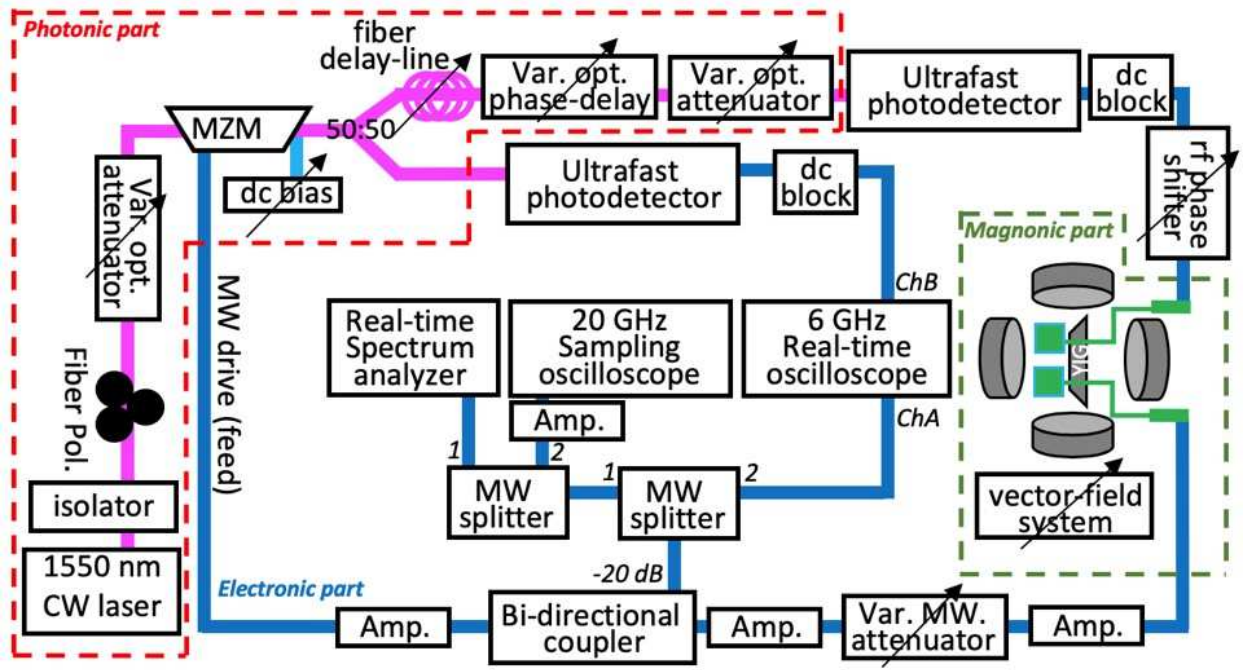}
 \caption{Schematic illustration of the MOEO experimental setup. The photonic part includes: CW 1550-nm laser source, fiber polarizer, variable optical attenuator, MZM, fiber splitter, bulky fiber delayline, variable fine phase-delay, and ultrafast photodetectors; the electronic part includes: dc-blocks, rf phase shifter, amplifiers, variable microwave (MW) attenuators, bi-directional coupler, MW splitters; the magnonic part includes: vector magnetic field system and a YIG spin wave delayline (replaceable by other magnonic delay structures). The data acquisition system part includes: a 6-GHz realtime oscilloscope for monitoring the low-frequency main harmonic modes of the OEO, a 20-GHz sampling oscilloscope for monitoring higher harmonics in the spin wave bands, and a realtime spectrum analyzer for monitoring frequency domain across a abroad bandwidth up to 26-GHz. Various bandpass filters are optionally inserted on-demand for different mode selections. Tunable-parameter components are indicated by an overlaid sloped arrow ($\nearrow$).}
 \label{fig1}
\end{figure*}

\begin{table*}[htb]
 \begin{tabular}{||l|l||} 
 \hline
  Spatially-distributed Parameters & Practical Control-elements \\ [0.8ex] 
 \hline\hline
  Main-line optical power & variable optical attenuator (mechanical or electrical tuning)    \\
 \hline
  Main-line optical phase (coarse delay) & bulky patch fibers  (for delay in the $\sim$ m or $\sim$km range)  \\
 \hline
  Main-line optical phase (fine delay) & picosecond phase shifter   (for delay in the $\sim$ ps or $\sim$ns range) \\
 \hline
  Feedback-line microwave power & variable microwave-attenuator  (electrical tuning, controllable steps-size) \\
 \hline
  Feedback-line modulation depth (MZM phase) & dc-bias voltage source  (electrical tuning, controllable steps-size) \\
 \hline
  Feedback-line microwave phase & microwave phase shifter  (mechanical tuning) \\
 \hline
  YIG spin-wave magnonic delay & magnetic field  (quadratic electromagnets, vector field) \\
 \hline
\end{tabular}
\caption{Summary of the complete, spatially-distributed parameter space of the MOEO system.}
\label{table1}
\end{table*}

%\begin{figure*}[htb]
 %\centering
 %\includegraphics[width=6.5 in]{Fig2.eps}
 %\caption{Photographs showing the software control and automation of the hardware for experimental parameter scanning: (a) central operation control, theTEC:OS, and examples of hardware modules used in the experiment, (b) variable attenuator (controls the \textit{G}), (c) dc-Bias (controls the \textit{V}), (d) spectra (frequency-domain) and (e) oscilloscope (time-domain), and (f) camera monitor. Detailed definition of the scanned parameters \textit{V, G, A, H} are in Sec. III.A. \textcolor{black}{More information about the module functionalities and the thaTEC:OS measurement protocol can be found at Ref[\cite{thatec}]. } }
% \label{fig2}
%\end{figure*}

\begin{figure*}[htb]
 \centering
 \includegraphics[width=5.8 in]{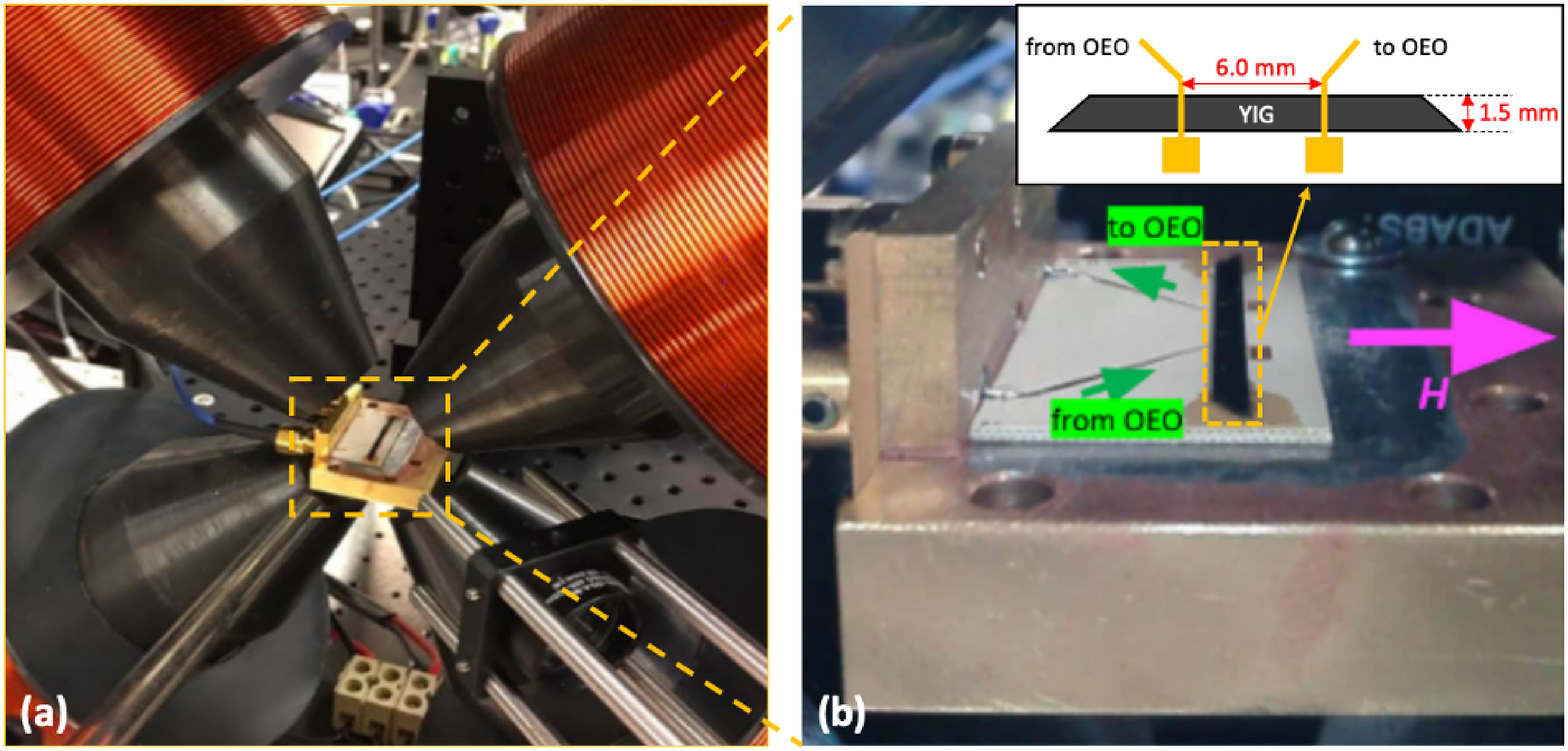}
 \caption{(a) Photographs of the key magnonic part consisting of (a) the vector magnet and (b) the YIG delayline. The magnetic field is vectorially constructed and is perpendicular to the YIG stripe, corresponding to the MSSW excitation configuration. \textcolor{black}{Inset of (b) shows the dimensions of the YIG trapezoid used as the magnonic delayline. }}
 \label{fig3}
\end{figure*}

While OEOs feature high quality(Q)-factor and phase coherence (this has been great in discrete rf-signal generations), nonlinear optical tuning is challenging to achieve in common systems. One solution is to interface with, in a hybrid but coherent fashion, other nonlinear mediums in an OEO system with external parameter controls. For example, recent research have focused on the interdisciplinary fields of microwave photonics, combined with the emerging fields of magnonics \cite{harder_ssp2018,huebl_prl2013,xufeng_prl2014,yili_prl2019,luqiao_prl2019,yili_jap2020}. Magnonics combine waves and magnetism and investigates the behavior of spin waves - magnons (in the language of quantum mechanics) - the collective excitation quanta of spin systems in magnetically-ordered materials \cite{chumak_nphys2015,chumak_jphysd2017,grundler_jphysd2010,slavin_ncomm2010}. Incorporating magnonic subsystems in conventional OEOs can immediately open additional degrees of freedom and enrich the nonlinear and chaotic dynamics in such a hybrid system. This is because of the inborn, nonlinear multi-magnon processes such as the two-magnon scattering \cite{platow_prb1998,heinrich_prb2004}, three-magnon process \cite{aaron_prl2009,lopes_pre1996,marsh_prb2012,camley_prb2014}, and four-wave mixing \cite{bloom_apl1977,serga_prl2005,celinski_apl2011,khitun_apl2008}. Magnonic systems can be more easily driven into nonlinear regimes due to their embedded nonlinearity in the equation of motion, i.e. the Landau–Lifshitz–Gilbert (LLG) dynamics, for spins in magnetically-ordered materials. Finally, magnons offer a great advantage since spin wave wavelengths are orders of magnitude smaller than the wavelengths of electromagnetic waves in the same GHz range, which allows for miniaturization of future oscillator devices.    

Earlier theoretical developments and pioneering proof-of-concept reports, especially from Kalinikos and colleagues \cite{vitko_jpcs2018,nikitin_emc2017,ustinov_metals2018,ustinov_magnlett2015,vitko_magnlett2018}, have shown the feasibility of such ``magnetically tunable" microwave optoelectronic oscillators.  However, a comprehensive tehnical examination of such a hybrid system is lacking. For example, a unique feature of such an integrated magnonic-opto-electronic oscillator (MOEO) system is the large yet spatially-distributed parameter space, which encloses photonics, electronics, and magnonics components. In this work, we demonstrate a comprehensive MOEO system with a high-quality Y$_3$Fe$_5$O$_{12}$ (YIG) magnetically-tunable delayline, and summarize the various technical aspects of such a system. We show a systematic and concurrent measurement protocol in the frequency-domain and time-domain, and reveal the rich nonlinear dynamics and combinatorial modes owing to the hybrid mixing of optoelectronic and spin wave modes. We discuss the complex spin wave dispersion patterns arising from the multiple tuning channels. Due to the ultrahigh sensitivity of the oscillator frequencies and harmonics to various external perturbations such as heat, light, and spin-torques, the MOEO system may find usefulness in spintronic-based signal generation, magnonic logic, sensing, and related applications.

\section{Construction of the magnonic-opto-electronic system}

Our MOEO system consists of photonic, microwave-electronic, and magnonic parts. The complete, spatially-distributed parameter space is illustrated in Fig. \ref{fig1} and also summarized in Table \ref{table1}. The software and automation control were achieved by using the thaTEC:OS automation platform (thaTEC Innovation GmbH) \cite{thatec}. Each individual set and read device are controlled by a dedicated module, which is then interfaced and programmed on-demand by the central operation system, thaTEC:OS.   

\subsubsection{Photonic part}

\textcolor{black}{As shown in Fig.\ref{fig1},} we use a continuous-wave (CW) 1550-nm laser source (Optilab LLC) with stabilized power output (fixed at $\sim$ 17 dBm). A fiber isolator is used after the laser to prevent possible reflected light and subsequent power disturbance. The fiber light polarization is then fine-adjusted to ensure a maximum power output before the variable optical attenuator, where the main-line optical power is tuned. The main-line optical power is also monitored by a in-line power-meter (Pure-Photonics). The modulation is achieved via a broadband, high-power MZM (Optilab LLC) modulating up to 12.5 GHz. The fiber optics then split the beam (50:50). One branch, i.e. the signal line, goes through an optional fiber delayline, fine phase delayline, and then another variable optical attenuator, before entering the ultrafast photodetector (up to 12.5 GHz, from EOT, Inc.). The delay can be controlled by adding bulk fibers (from 1 m to $\sim$ 3 km) for coarse tuning or by using a variable fiber delay-line ($<$ 600 ps) for fine tuning. The other optical line serves as a main reference line for the lower-GHz harmonics which are directly monitored by a 6-GHz real-time oscilloscope (Keysight Technologies).\\

\subsubsection{Electronic part}

The photodetector then converts the ultrafast optical signal to a microwave signal, which subsequently travels through a set of microwave components. A dc-block is placed after the photodetector to remove any dc background possibly incurred at the photonic part. In a generic OEO system without the magnonic part, the signal passes through a tunable rf phase shifter, a series of amplifiers and a variable attenuator, before sent back to the MZM to close the feedback loop. Along the way, a 20-dB microwave coupler is used to extract a small portion of the signal for realtime monitoring and analysis. We use a broadband realtime spectrum analyzer (up to 26-GHz, from Berkeley Nucleonics Corp.) for frequency-domain and spectrogram representation. For time-domain study, besides the 6-GHz realtime oscilloscope primarily for monitoring the fundamental OEO auto-oscillating modes (higher-MHz to lower-GHz), another 20-GHz sampling oscilloscope (Pico Technology) is inserted to monitor the higher harmonic modes as well as the interfering wave packets and chaotic waveforms, especially in the spin wave bands (1 - 10 GHz) after incorporating the magnonic part as will be discussed below.\\

\subsubsection{Magnonic part}

\begin{figure*}[htb]
 \centering
 \includegraphics[width=7 in]{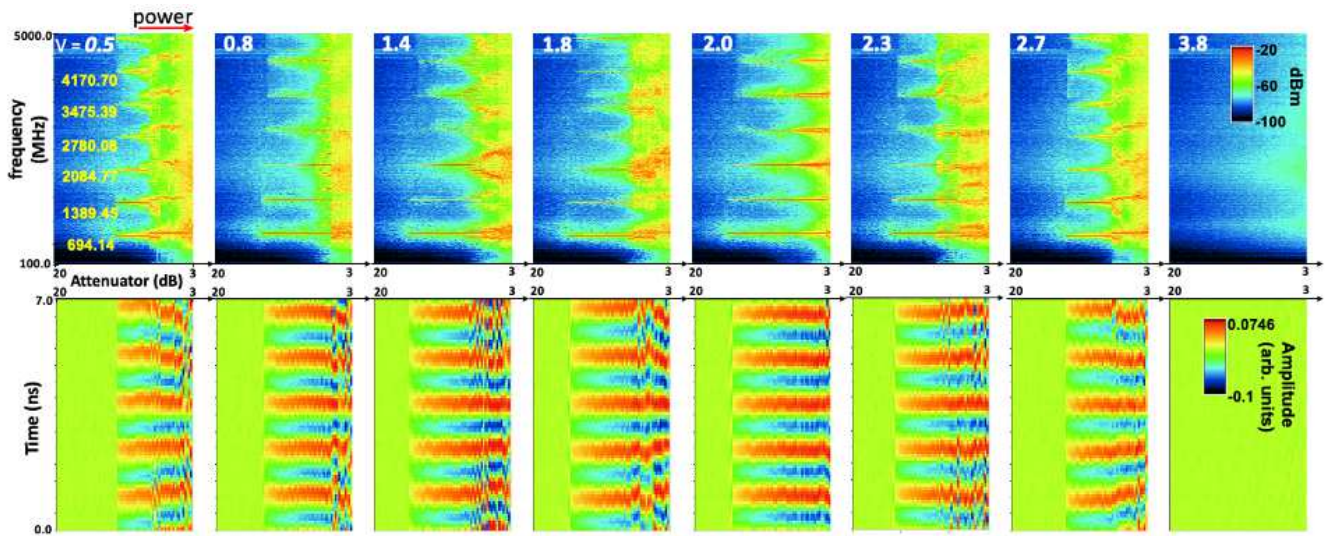}
 \caption{The spectrum (top panel) and waveform (bottom panel) of the microwave signals as a function of the ring gain and at different MZM dc Bias values. Harmonic frequencies are observed with a fundamental frequency at 694.1 MHz and $\Delta f \sim$ 695 MHz. \textcolor{black}{The horizontal axis is the power attenuation value in dB, scanned from 20 to 3 dB. The smaller the attenuation, the higher power flowing inside the loop. }}
 \label{fig4}
\end{figure*}

The magnonic part consists of a Y$_3$Fe$_5$O$_{12}$ (YIG) delayline made by liquid-phase epitaxy on Gd$_3$Ga$_5$O$_{12}$ (GGG) substrate and cut to a long, narrow film strip, see Fig. \ref{fig3}. The thickness of the YIG is 7.8 $\mu m$. The two ends of the YIG film strip are tapered to a trapezoid to minimize spin wave reflection, \textcolor{black}{see Fig.\ref{fig3}(b)}. The width of the delayline is 1.5 mm. Two microstrip transducers on the YIG strip serve as the input (launching) and output (receiving) antennas, respectively. The distance between the antennas is 6.0 mm. The antenna converts the power of the electromagnetic waves into spin waves and vice versa, and the rf field from the antenna can be large enough to excite nonlinear spin wave modes as the loop gain is sufficiently large. The spin wave propagation inside the YIG is governed by the dispersion relationship\cite{kreisel2009microscopic, lim2019direct}. \textcolor{black}{For our experimental configuration, the magnetic field is perpendicular to the wavevectors of the propagating spin waves, corresponding to the magnetostatic surface spin waves (MSSWs). Under the excitation geometry of such surface spin waves,} the YIG can be viewed as a highpass filter (or a non-ideal bandpass filter) whose low-frequency threshold is cut at the FMR frequency ($k = 0$). Therefore, the new harmonic frequencies ($k >0$) generated via various nonlinear spin wave processes can be received and converted by the output antenna that subsequently launches post-generation spin waves back to the YIG, or circulates (in electromagnetic wave form) through the photonic part and re-enters the input antenna. Such a property is critical to the observations of harmonic frequencies to be discussed later. The strength of the applied magnetic field is another critical factor that not only serves as a requisite condition for the self-generation in the MOEO, but also is needed for the onset of the nonlinear interactions and the excitation of chaos. In our setup, we use a quadratic-pole electromagnet system supplying vector magnetic fields up to 0.38 T, which conveniently enables the various configurations for spin wave activation, such as surface or volume waves on-demand (details to be discussed later). Overall, the YIG delayline represents a broadband, passive device with continuous frequency spectrum, tunable time delay, and a host of nonlinear behaviors, whose characteristics can be readily controlled by the external magnetic field besides microwave power.\\

%\textcolor{black}{[VNA characterizations will be added here soon for assisting analysis of the spin wave group velocity and the delay time.]}   

\section{Results and Discussions}

\subsection{Generic Optoelectronic Performance}

We first focus on the generic signal feedback behaviors of the OEO subsystems without inserting the YIG delayline. In Table \ref{table1}, we have outlined the parameter space, in which the most important controls are the MZM dc-bias, \textit{V}, the feedback gain, \textit{G}, the external magnetic field, \textit{H}, and the delay time, \textit{$\tau$}. The feedback gain is usually defined as the onset power level for observing the auto-oscillation. Here, it is continuously tuned by the attenuation level, \textit{A}, of the loop via a variable microwave attenuator. Earlier works using a YIG feedback loop were conducted under either a fixed \textit{V} \cite{ustinov_metals2018,ustinov_magnlett2015,vitko_magnlett2018} or \textit{H} \cite{watt_prap2020}, thus the \textit{G} is a fixed number throughout the discussion. Here, due to the different auto-oscillation thresholds of the feedback loop with respect to changing \textit{V,H}, the gain will have to be defined differently for each scenario. We thus find it is more convenient to use just the attenuation level, \textit{A}, throughout the discussion. \textit{V, A, and H} will be scanned with high resolution over a broad range via automation, and the delay time will be manually varied by adjusting the optical fiber length, the phase shifter, and the spin waves in YIG.

Figure \ref{fig4} shows the OEO signals (without inserting the magnonic part) in the frequency (top panel) and time (bottom panel) domains. The spectra and realtime oscillations in each subplot are scanned with increasing the loop gain (reducing the attenuation level, from 20 to 3 dB, of the variable attenuator). Harmonic frequencies are sharply generated at a certain attenuation level, which can be defined as the reference gain, i.e. \textit{G} = 0 dB, to indicate the onset of the auto-oscillation. We observe a series of the signal harmonics at 694.1, 1389.5, 2084.8... MHz, which corresponds to a short time period of 1.4 ns \textcolor{black}{(Fig. \ref{fig4}, top panel)}. As the ring gain increases, the signal starts to bifurcate and eventually enters the chaotic regime. The thresholds for the auto-oscillation are slightly different at different MZM dc bias within the operational window, as seen in Fig. \ref{fig4}, but otherwise the observed feature are all similar. Such a dependence on the dc Bias is also easily seen by the scanned (\textit{V, A}) phase diagram in Fig. \ref{fig5}, which we will discuss further below and in comparison with the insertion of the magnonic (YIG) part.

\subsection{Magnonics-based mode-filtering}

The use of the vector magnet allows us to construct a 2D magnetic field via vector addition, see Fig. \ref{fig3}. Generally, spin waves in the YIG strip can be excited in three different configurations with distinct dynamic properties, i.e., forward volume spin wave, backward volume spin wave, and magnetostatic surface spin wave (MSSW)\cite{damon1961magnetostatic, lim2019magnetostatic, lim2018forward}. \textcolor{black}{In this present work,} as an example, we use MSSWs for our MOEO study and demonstrate auto-oscillations, mode selecting and filtering, as well as harmonic modes intermixing. \textcolor{black}{However, we note that the system, in general, could also be used to investigate other types of spin wave and/or spin dynamics involving complex field-scan sequences. } 

In the MSSW configuration, the external magnetic field \textit{H} is applied in the plane of the YIG, and perpendicular to the direction of spin wave propagation. We measure the transmission characteristics (via the microwave coupler) by scanning the dc bias and the gain, and therefore construct the 2D phase-diagram of (\textit{V, A}). 

\begin{figure}[htb]
 \centering
 \includegraphics[width=3.4 in]{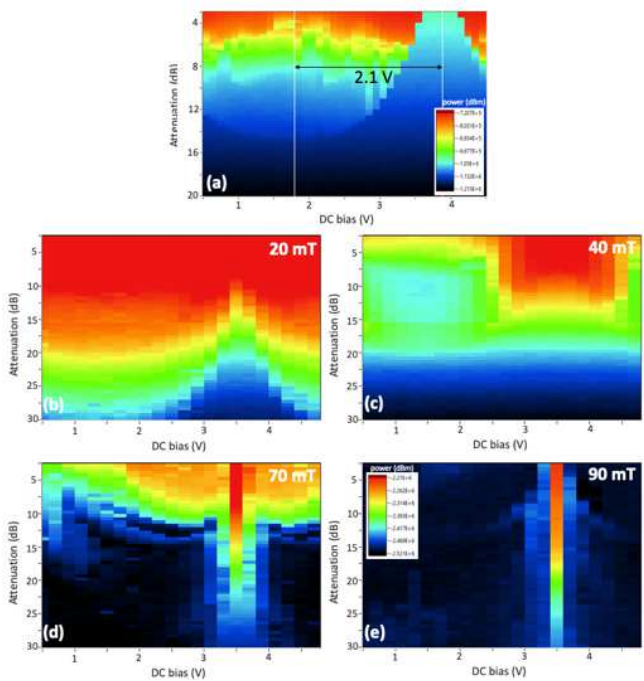}
 \caption{Scanned (\textit{V, A}) phase diagrams, (a) without the YIG delayline, and with the YIG at different magnetic fields, (b) 20 mT, (c) 40 mT, (d) 70 mT, and (e) 90 mT. }
 \label{fig5}
\end{figure}

Figure \ref{fig5}(a) first shows such a 2D map for a measurement without the magnonic part. The transmission window lies from 0 V to $\sim$ 3.9 V and peaks at $\sim$ 1.8 V. The minimum transmission point is 2.1 V away from the maximum, corresponding to the $\frac{1}{2} V_\pi$ of the MZM, which are expected for typical OEO performance. After inserting the magnonic part, the YIG component simultaneously plays the role of a mode filtering, time delay, and nonlinear element. Due to the spin wave mode selection, a significant amount of power transmission is reduced. While the phase diagram by itself largely represents generic characteristics of the electronics of the OEO loop, the magnetic field dependence reflects the selective power transmission due to magnonic mode selection in the YIG strip, and is clearly evident from our measurements.

Figure \ref{fig5}(b)-(e) compares the (\textit{V, A}) diagrams of power transmission in the presence of the YIG delayline at different applied magnetic fields. The transmitted power is an order of magnitude smaller than the case without the YIG in Fig. \ref{fig3}(a). At \textit{H} = 20 mT [Fig. \ref{fig5}(b)], the transmission across the \textit{V} range becomes more continuous than in (a), where the peak value can still be recognized (at lower gain values), around 1.8 V. Then, increasing the field to \textit{H} = 40 mT [Fig. \ref{fig5}(c)] results in a strong shift of the transmission pattern, in which a strong power transmission (of spin waves) emerges around the $V_\pi$ point of the MZM. As the field further increases, the transmission window keeps narrowing down at 70 mT, Fig. \ref{fig5}(d), and eventually to a very narrow transmission regime, once the magnetic field passes the YIG spin wave bands around 90 mT, in Fig. \ref{fig5}(e). The (\textit{V, A}) phase diagram provides an overall evaluation of the change in the total power transmission pattern of the microwave signals by the insertion of the YIG delayline. It is desirable to gain a closer look at the details of the spin wave bands upon scanning the parameters \textit{V}, \textit{A}, and \textit{H}.

\subsection{Magnetic-field-tunable signal channels}

One of the initial motivations in studying spin wave propagation was the realization of microwave filters for analog data processing. For example, studies of magnonic crystals demonstrate engineered spin wave spectra which comprise regions of frequencies for which the transmission of the microwave signals is prohibited, like a bandgap \cite{chumak_nphys2015,chumak_jphysd2017,grundler_jphysd2010,slavin_ncomm2010}. The central frequencies of the pass/stop bands can be readily controlled by the applied magnetic field. However, to-date there has been no report showing the use of magnonic crystals in auto-oscillating systems. 

\begin{figure}[b]
 \centering
 \includegraphics[width=3.4 in]{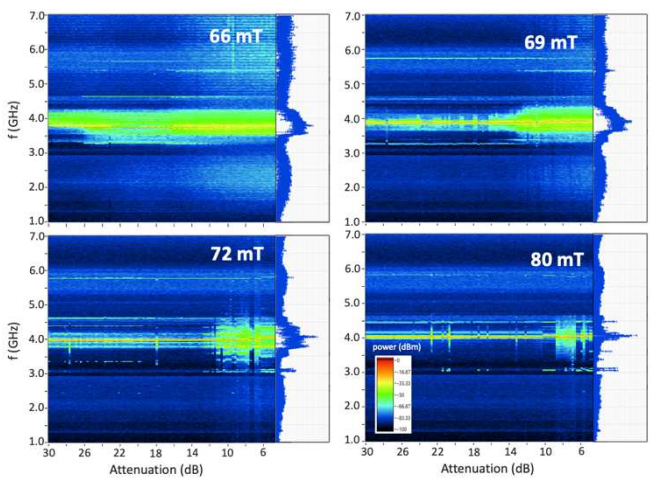}
 \caption{Evolution of the spin wave bands versus the ring attenuation at different external magnetic field near a $\sim$ 4 GHz OEO mode. The right panel is the example 1D trace taken at a ring attenuation level of 3.75 dB. }
 \label{fig6}
\end{figure}

Here, one important advantage of the MOEO lies in the existence of two-tone excitation that are of very different nature: one is the OEO photonic mode that features sharp resonance profile and high spectra purity, and is relatively insensitive to the magnetic field; the other is the magnonic mode that is significantly more lossy, and susceptible to magnetic fields and nonlinear effects. Different from a pure YIG feedback ring where the fundamental mode (and all the higher harmonics modes) are both of magnetic origin (Kittel modes), the MOEO allows tuning the Kittel modes close to or away from the optoelectronic (OEO modes) by changing magnetic field. Figure 6 shows an example of an OEO harmonic mode tuned near a spin wave band \textcolor{black}{(at $\sim$ 4 GHz) and at selective field steps, from 66 mT to 80 mT}. At 66 mT, a broad excitation band covering from $\sim$ 3.3 - 4.2 GHz are found, whose amplitude persists even at quite high attenuation \textit{A} (low gain) level. At 69 mT, the spectrum significantly narrows down at high \textit{A} values (above $\sim$ 15 dB), demonstrating a magnetic-field-tuned bandpassing. As the field further increases to 72 mT, the passband continues to narrow for \textit{A} $>$ 12 dB, and in addition, the higher power regime also starts to exhibit some instability over the power level. Finally, at 80 mT, only a very narrow spectra near the OEO mode can be seen at almost the whole power attenuation range, which is accompanied also by a significant instability at higher power levels (\textit{A} $<$ 7 dB). Finally, the broad spin wave bands is due to the thick \textcolor{black}{and wide} YIG strip that hosts a broad a range of wavenumbers and also due to the perpendicular standing spin wave (PSSW) modes that are co-excited which will be discussed later. \textcolor{black}{Using thinner and narrower YIG samples and/or magnonic crystals would potentially result in a narrower passband and also mitigate the PSSW mode excitations. \cite{yili_YIGPy,xiong_srep2020}}

\subsection{Combinatorial time delay}

The total delay time is an important parameter to the frequency self-generation and memory effect in the auto-oscillating system \cite{vasyl_srep2014}. Generally, frequency generation enters after a positive delayed feedback has been added to the system. If the loop gain is sufficient to compensate the losses of the loop, a signal can circulate in the ring, thus forming a ring resonator with a discrete set of frequencies. In opto-electronic systems, the time delay is achieved by light traveling in the photonic fibers or other optical phase-changing components \cite{yanne_rmp2019} \textcolor{black}{(photonic and electronic parts)}. In the active ring configuration, the microwave signal from the output antenna is amplified and then injected back to the input end \cite{wu_ssp2010}. Delay components such as rf phase shifter and/or electronic delay \cite{riou_prap2019} can be inserted in the microwave path. The time delay in the YIG strip is achieved by spin waves traveling between the two antennas, resulting in a phase shift of the microwave signal across the YIG strip \textcolor{black}{(magnonic part)}. Therefore, the total time delay of the loop is: $\tau = \tau_p + \tau_e + \tau_m$, in which the subscripts, $p, e, m$ represents photonic, electronic, and magnonic delays. Correspondingly, the resulting phase shift can be written as:$\Phi = \Phi_p + \Phi_e + \Phi_m$ \cite{ustinov_magnlett2015}. Here, $\Phi_p = n_{eff}Kl_{p}$ is the linear optical phase shift, where $n_{eff}$ is an effective refractive index in the fiber, $K$ is the optical wavenumber, and $l_p$ is the length of the fiber. $\Phi_e$ is the electronic phase that accounts for the microwave cables, amplifers, and rf phase shifter. $\Phi_m = kl_m$, in which $k$ is the spin wave wavenumber and $l_m$ is the length of the delay line. Often times it is also necessary to include the nonlinear counterparts of the phase contributions from the above components, in which the spin wave nonlinear contribution to phase, $\Phi_m^{NL} = - N \frac{m^2}{2M_0} l_m / v_g$, would play a nontrivial role such as the microwave bistability \cite{vitko_magnlett2018}. In the formula, $N$ is the nonlinear self-coefficient for MSSW, $m$ is the transverse magnetization, $M_0$ is the total magnetization, and $v_g$ is the spin wave group velocity.    

In our measurements, we have tested our MOEO system with different fiber lengths ranging from short delays (1, 2, 5 m) and long delays (500, 1000, 2000 m). The resultant time delay ranges from several ns to $\sim$ 5 $\mu$s. Previous works have mostly focused on long delay from the photonic part. Here we explore the situation where the two delay time are comparable and also where the magnonic delay is even more than the optical delay. Overall, we found that the spectral purity is largely enhanced at longer delay time by the elimination of higher-order harmonic modes, however, the dominant OEO mode and its intermixing with the Kittel mode are always observed (to be discussed later). In addition, electronic delay such as using microwave cables and the rf phase shifter yields similar results that are not interesting within the present context.  However, we note that the rf shifter is found to be very effective in fine-tuning the OEO modes owing to the larger photon wavelength at the microwave frequencies. On the other hand, the magnonic delay due to the YIG strip is governed by the relevant propagating spin waves and their dispersion relationship and group velocities, which we will separately discuss in detail in below.

\subsection{Spin wave dispersion and group velocity}

\begin{figure*}[htb]
 \centering
 \includegraphics[width=6.8 in]{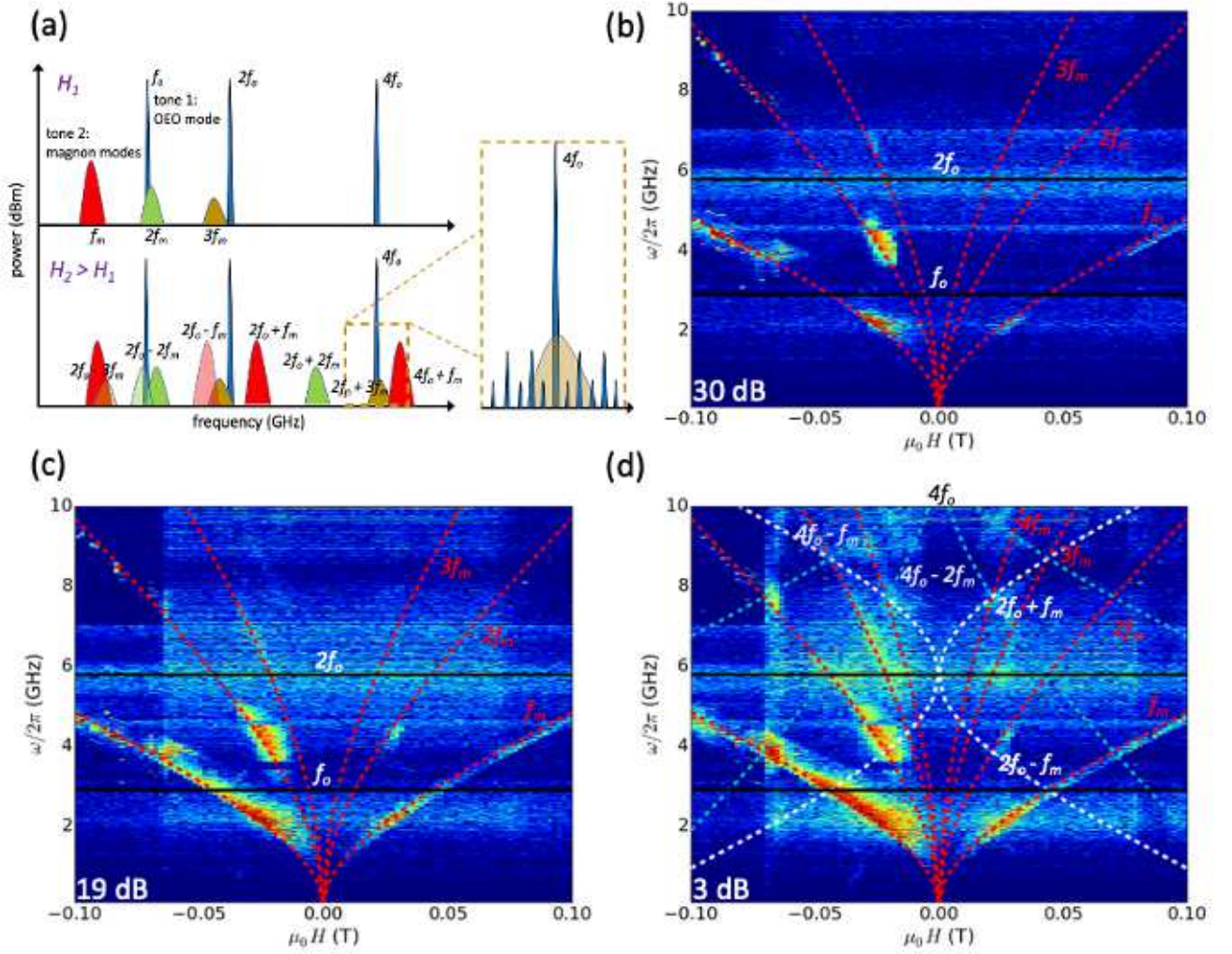}
 \caption{(a) Schematic illustration showing the two-tone excitation from the OEO modes, $f_o$ and its harmonics, and the Kittel modes, $f_m$. Harmonic beating modes are generated due to nonlinear excitation, which extends the signal self-generation to regions further away from the dominant OEO modes by the external field. The zoom-in section illustrates the four-wave mixing process when the spin wave band span across the the OEO mode. Example dispersion maps of the MOEO modes measured at different attenuation levels, (b) 30 dB, (c) 19 dB, and (d) 3 dB. }
 \label{fig7}
\end{figure*}

In the YIG delayline, which eigenmodes are excited depends on the frequency characteristics, i.e., the delayline's transduction frequency band and the spin wave dispersion relation. Fig. \ref{fig7} plots the MOEO spectra as a function of the applied magnetic field. Figure \ref{fig7}(a) summarizes the OEO ($f_o$) and Kittel ($f_m$) modes and the process of harmonic modes generation with respect to the magnetic field. Fig. \ref{fig7}(b)-(d) shows the example dispersion maps at different attenuation levels, 30, 19, and 3 dB. The dashed lines are corresponding fits to the harmonic spin wave beating modes. 

\begin{figure}[b]
 \centering
 \includegraphics[width=3.4 in]{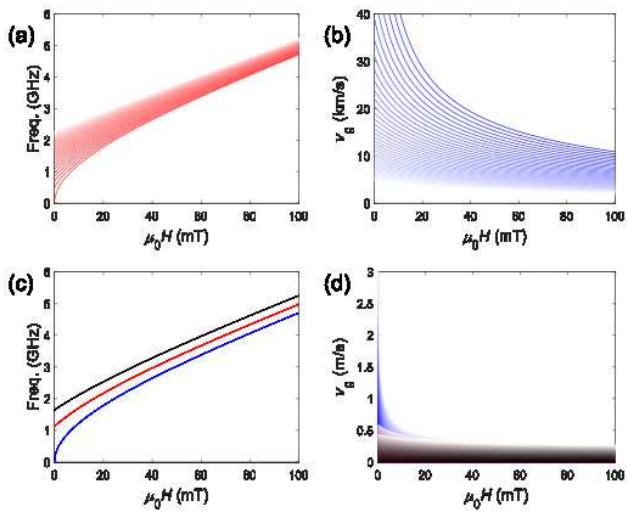}
 \caption{(a) Calculated MSSW frequency band as a function of $\mu_0 H$. The plotted $k$ range is chosen from 0 to $10^5$ rad/m to match the experimental observation. (b) The corresponding MSSW group velocity bands as a function of $\mu_0 H$ (lighter colors indicate larger $k$ values). (c) Calculated PSSW frequencies as a function of $\mu_0 H$, showing example trace of $n$ = 1 (blue), $n$ = 35 (red), and $n$ = 50 (black). The plotted $k$ range is chosen from 0 to $10^5$ rad/m. (d) PSSW group velocity bands as a function of $\mu_0 H$, showing the cases for $n$ = 1 (blue), $n$ = 35 (red), and $n$ = 50 (black). }
 \label{fig8}
\end{figure}

The magnetic field direction supports the MSSW modes.  First, at frequencies below $\sim$ 5 GHz we observe two strong branches with symmetric dispersion curve for positive and negative fields, Fig. \ref{fig7}(b). They correspond to the dominant MSSW modes that can be expressed as \cite{kalinikos_jpc1986}:
\begin{widetext} 
\begin{equation}
f_{MSSW}(H, k) = \sqrt{(f_H + \lambda_{ex}^2 f_M k^2)(f_H + \lambda_{ex}^2 f_M k^2 + f_M)+\frac{f_M^2}{4}(1-e^{-2kd})}
\label{eq01}
\end{equation}
\end{widetext} 
where $f_H = \gamma H$, $f_M = \gamma (4 \pi M_s$), $H$ is the applied magnetic field, $M_s$ is the saturation magnetization, $d$ is the thickness of the YIG, $k$ is the wavevector, and $\gamma$ is the gyromagnetic ratio, $\lambda_{ex}$ is the exchange length and is equal to $\sqrt{A_{ex}/2 \pi M_s^2}$, and $A_{ex}$ is the exchange stiffness. In Fig. \ref{fig7}(b,c,d), we also show the fitting curve by taking the dominant OEO mode as $f_o = \omega_{o}/2\pi=2.9$ GHz, with the mode beating taking place for both $\omega_{o}$, $2\omega_{o}$, and $4\omega_{o}$. We take the values for YIG, $\gamma/2\pi$ = 28 GHz/T, $\mu_0 M_s$ = 0.1835 T, $A_{ex}$ = 3.5 pJ/m, and \textit{d} = 7.8 $\mu m$, in our analysis. The fitting curves nicely reproduce the experimental spectra. The MSSW group velocities, $v_{g_M}$, of spin wave in the YIG delayline can be derived as: 
\begin{widetext} 
\begin{equation}
v_{g_M} = \frac{\partial f}{\partial k} = \frac{2 \lambda_{ex}^2 f_M k (f_H + \lambda_{ex}^2 f_M k^2 + f_M/2) + f_M^2 d e^{-2kd}/4}{\sqrt{(f_H + \lambda_{ex}^2 f_M k^2)(f_H + \lambda_{ex}^2 f_M k^2 + f_M)+\frac{f_M^2}{4}(1-e^{-2kd})}}
\label{eq02}
\end{equation}
\end{widetext}
As the attenuation is decreased, the auto-oscillation spectrum becomes stronger and broader in frequency, especially towards higher frequency, as seen in Fig. \ref{fig7}(c). This is due to the onset of auto-oscillation for the MSSW modes with higher $k$, which require larger gain to compensate the energy loss for reaching self-oscillation. Considering the range of the spin wave modes excited in our experiment as shown in Fig. \ref{fig7}, we calculate the spin wave dispersion and group velocities versus the magnetic field as shown in Fig. \ref{fig8}(a) and (b). The spin wavevectors range from 0 to $10^5$ rad/m. The spin wave velocities can range from $\sim$ 10 to 30 km/s, corresponding to a delay time from $\sim$ 200 to 600 ns, at $H$ = 20 mT for example.

Besides the two main MSSW modes, many additional auto-oscillation branches appear further away from the dominant MSSW waveband when the attenuation is decreased, as can be seen in Fig. \ref{fig7}(d). These new modes are the higher-order harmonics of the MSSW modes intermixed with the OEO modes that are nearly independent of $H$. In addition, we also observe auto-oscillation branches with frequencies that decrease with $H$ (beating modes), especially for 3 dB attenuation.  In Fig. \ref{fig7}(d), up to the 4$^{th}$ order harmonics of the MSSW modes can be observed. Both the higher harmonics and the intermixed modes are due to the nonlinear effects of the magnetization dynamics in the YIG \cite{slavin_ieee2009,muduli_PRB2010,ebels_APL2014}, especially in the auto-oscillation regime where the dynamics is of larger amplitude. 

In addition, the broadening of the auto-oscillation spectra may also come from the excitation of perpendicular standing spin wave (PSSW) modes along the thickness direction. Since the thickness of the YIG delay line is 7.8 $\mu$m, the PSSW modes are highly degenerate and can mix with the MSSW modes with higher in-plane $k$ values \cite{grundler_prap2019,xiong_srep2020}. The dispersion relation for the YIG PSSWs are written as \cite{demokritov_physrep2001,gui_prl2007}: 
\begin{widetext} 
\begin{equation}
f_{PSSW}(H, k) = \sqrt{[f_H + \lambda_{ex}^2 f_M k_{in}^2 + \lambda_{ex}^2 f_M (\frac{n \pi}{d})^2] [f_H + \lambda_{ex}^2 f_M k_{in}^2 + \lambda_{ex}^2 f_M (\frac{n \pi}{d})^2 + f_M]}
\label{eq03}
\end{equation}
\end{widetext}
where the total wavenumber $k_{tot} = \sqrt{k_{in}^2 + k_{perp}^2}$, $k_{in}$ is the in-plane wavenumber, and $k_{perp} = n \pi /d$, $n$ is the PSSW mode number. The available PSSW modes and their wavevectors are constrained by the film thickness. Due to the relatively thick YIG film in our present study, $d$ = 7.8 $\mu m$, the frequency gap between adjacent PSSW modes are quite close to each other, appearing as a spin wave band especially at higher $n$ values \cite{xiong_srep2020}. The PSSW spin wave group velocities, $v_{g_P}$, can be given as: 
\begin{widetext} 
\begin{equation}
v_{g_P} = \frac{\partial f}{\partial k} = \frac{2 \lambda_{ex}^2 f_M k_{in} [f_H + \lambda_{ex}^2 f_M k_{in}^2 + \lambda_{ex}^2 f_M (\frac{n \pi}{d})^2] + f_M/2}{[f_H + \lambda_{ex}^2 f_M k_{in}^2 + \lambda_{ex}^2 f_M (\frac{n \pi}{d})^2] [f_H + \lambda_{ex}^2 f_M k_{in}^2 + \lambda_{ex}^2 f_M (\frac{n \pi}{d})^2 + f_M]}
\label{eq04}
\end{equation}
\end{widetext}
The PSSW spin wave dispersion and group velocities versus the magnetic field are shown in Fig. \ref{fig8}(c) and (d), at example $n$ values, 1, 35, and 50. Compared with the MSSWs, the PSSW spin wave velocities are only a few m/s, thus they are expected to play a negligible role in the spin wave propagation of the YIG delayline even if they are present in the dispersion curves.  

\subsection{Optoelectronic and magnonic beating modes}

We then \textcolor{black}{come back to Fig.\ref{fig7} and} concentrate on the resonance harmonics and the nonlinear mixing of spin waves under the two tone excitation unique to our MOEO. It is noted that conventional optical fiber nonlinearity are difficult to achieve; and, the advantage of using a YIG delayline is its narrow linewidth, which allows, even at modest applied microwave power, the access to the nonlinear responses with abundant features in high-order dynamics including phase conjugation (time reversal), two-magnon, three-magnon process, and four-wave mixing and squeezing \cite{platow_prb1998,heinrich_prb2004,aaron_prl2009,lopes_pre1996,marsh_prb2012,camley_prb2014, bloom_apl1977,serga_prl2005,celinski_apl2011,khitun_apl2008}. Experimental methods towards the nonlinear effects also often involve parametric pumping, spin-wave bullets (soliton), and dynamic spin-wave channels \cite{serga_prl2005,melkov_jap2005,melkov_prb2009}. Only recently, four-wave mixing and phase conjugated magnons are reported in the continuous wave (CW) regime by using spatially separated pump and probe beams \cite{inglis_jmmm2019}. However, current reports involving two-tone auto-oscillations and nonlinear harmonics are only contained within the context of pure magnon modes \cite{bloom_apl1977,serga_prl2005,celinski_apl2011,khitun_apl2008,inglis_jmmm2019}.     

\begin{figure*}[htb]
 \centering
 \includegraphics[width=6.7 in]{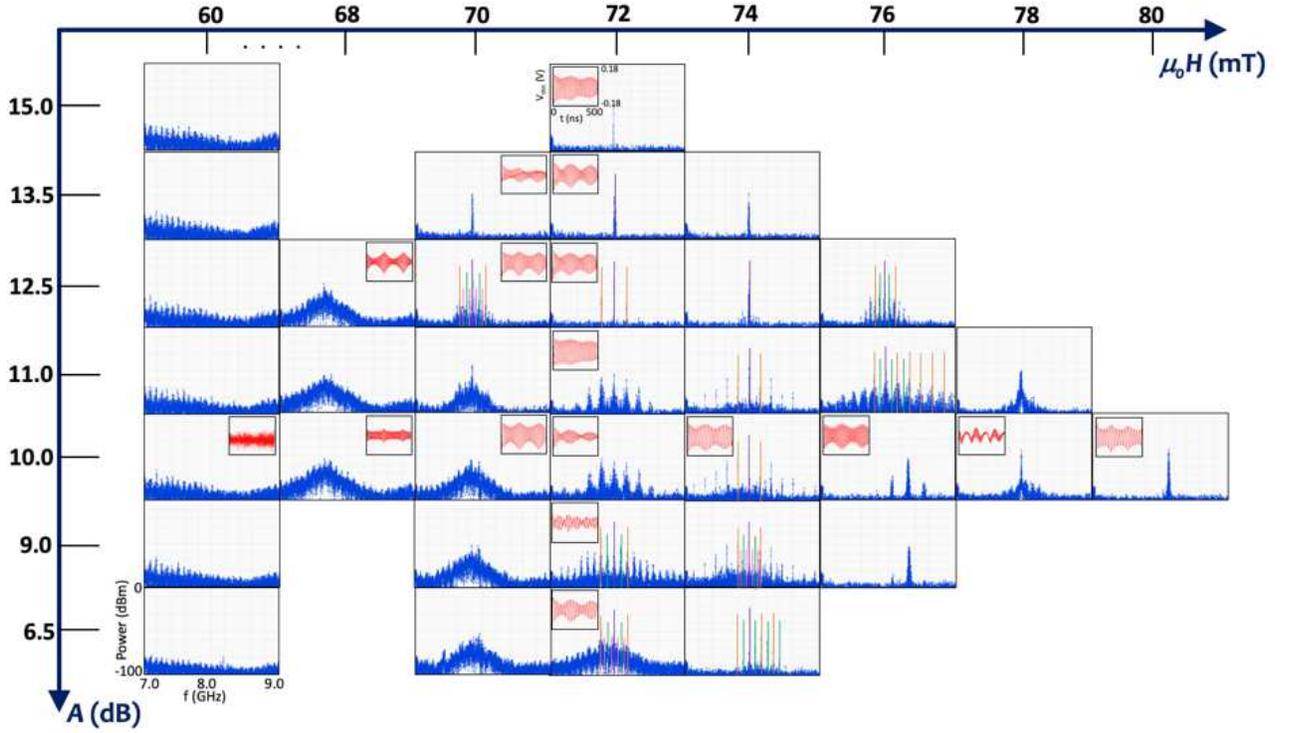}
 \caption{Evolution of the frequency- and time-domain spectra for a small magnetic field window (scanned from 60 to 80 mT) and at different loop attenuation level (from 15 to 6.5 dB). For the frequency spectra, the frequency range of interest is from 7.0 to 9.0 GHz, to enclose an local OEO harmonic mode center around $\sim$ 8 GHz, and the power is scanned from -100 to 0 dBm. For the time spectra, the scanned time window is 0 to 500 ns. Only selective 1D traces are shown due to their representative features. The harmonics from different generations are marked by thin vertical lines in the plot.}
 \label{fig9}
\end{figure*}

We can understand the origin of the mode beating effect by examining the macrospin version of the Landau-Lifshitz equation: 
\begin{equation}
\frac{d\mathbf{m}}{dt} = - |\gamma| \mathbf{m} \times \mathbf{H_{eff}} . 
\label{eq05}
\end{equation}
The transverse dynamic magnetization, $m$, couples to the rf field, \textbf{h}, from the antennas. Due to the delayline structure with the two transducers, the nonlinear excitation at both ends of the delayline introduce a two-tone excitation as illustrated in Fig. \ref{fig7}(a). One tone is the OEO mode (whose fundamental frequency is at $f_o$) with the rf field amplitude, $\mathbf{h_o}$. This mode is strong, narrowband, and sensitive to the accumulated phase of the whole circuit and be effectively fined-tuned by the rf phase shifter. The other tone, as introduced earlier, is due to the excitation of the Kittel mode, $f_m$, from the YIG, with an excitation amplitude, $\mathbf{h_m}$. The beating modes observed in Fig.\ref{fig7}(a) result from the nonlinear interactions between the OEO tone and the propagating magnonic tone in the YIG delayline, similar to the observations in spin-torque oscillators. The antenna converts the power of the electromagnetic waves into  spin  waves  and  vice versa, and, new  harmonic frequencies  generated via four-wave processes  can  be also received  and  converted  by one antenna that subsequently launches post-generation spin waves in the YIG after circulating in the loop and re-entering the other antenna. Therefore, the total rf excitation field, \textbf{h}, under the self-generation, has the dual components:
\begin{equation}
\mathbf{h} = h_o e^{i \omega_o t} + h_o^* e^{-i \omega_o t} + h_m e^{i \omega_m t} + h_m^* e^{-i \omega_m t}
\label{eq06}
\end{equation} 
where $\omega_o = 2 \pi f_o$ and $\omega_m = 2 \pi f_m$ are the angular frequencies for the OEO and Kittel tones, respectively. Under high power and in the nonlinear regime, the transverse magnetization are given by considering $h$ and the susceptability $\chi$ with including also the higher odd-order terms (the even orders are omitted due to its thermal origin): $\mathbf{m} = \mathbf{\chi^{(1)} h} + \mathbf{\chi^{(3)} h^3 + ...}$. The higher order terms allow introducing the harmonics terms, for example, the higher magnon harmonics, $h_m^3 e^{i (3\omega_m)t}$, $h_m^{*3} e^{-i (3\omega_m)t}$, as well as the mixed spin wave and OEO modes, $h_m h_o^2 e^{i (2\omega_o + \omega_m)t}$ and $h_m^* h_o^2 e^{i (2\omega_o - \omega_m)t}$, and so on. In our present work, we observe the two-tone beating modes involving up to the 4-th of OEO harmonics and the 3-rd of spin wave harmonics. 

A better insight may be gained by evaluating the spectral evolution when the spin wave modes are tuned (by the magnetic field) to intersect with an OEO mode and at different loop attenuation levels. Figure \ref{fig9} presents the evolution of the frequency spectra for a small magnetic field window (scanned from 60 to 80 mT) and at different loop attenuation level (from 15 to 6.5 dB). The frequency range of interest is from 7.0 to 9.0 GHz, to enclose an local OEO harmonic mode center around $\sim$ 8 GHz, and the power is scanned from -100 to 0 dBm. Only selective 1D traces are shown due to their representative features as discussed below.  

\textcolor{black}{In Figure \ref{fig9},} at 60 mT, before the spin wave band kicks in, no strong harmonics are observed even for the OEO modes. This reflects the property that YIG delayline being a good narrowband rf filter similar to previously reported \cite{chumak_jphysd2017,ustinov_magnlett2015}. As field increases to 68 mT, a wide spin wave band shows up, and then at 70 mT, distinct OEO harmonics can be observed at selective attenuation level, e.g., 13.5 to 10 dB. As the spin wave band further shifts to the higher frequency, at 72 mT, the OEO harmonics can be more broadly observed for almost the whole power range. In particular, by tracing the plots vertically, the power evolution of the spectra at this magnetic field shows a process of frequency-halving (periodic doubling) due to four-wave mixing. First, a single, strong OEO mode is observed at $A >$ 13.5 dB, and two sidebands emerges at $A =$ 12.5 dB. Second, further reducing the loop attenuation results in more sidebands and also the broadening of the mode spectra. These modes are uniformly spaced and centered around the dominant mode. Then, at $A =$ 9.0 dB, a clear frequency-halving (periodic doubling) due to the four-wave process is evident. Finally, the spectrum further broadens at $A =$ 6.5 dB. The harmonics from different generations are marked by thin vertical lines in the plot. The corresponding time traces measured concurrently by the 20-GHz oscilloscope is also included as the insets to each key frequency spectrum. 

Similar features are also present for $H =$ 74 and 76 mT, however, as the spin wave band shifts to higher frequency with the magnetic field, the dominant harmonic mode and sidebands also shift to the right. Eventually, as the spin wave band moves out of the frequency window, for example at 80 mT, the OEO modes are also greatly attenuated. The OEO harmonics are known to hold important applications in optoelectronics such as low-noise rf generation, frequency combs, and signal amplification. The four-wave mixing is also an important process towards nonlinear signal generations, in its close context with the modulation instability and chaotic spin wave excitation. The OEO modes are usually set at discrete frequencies locked by the loop delaytime. On the other hand, the Kittel modes are smoothly tuned by the magnetic field, and the nonlinear mixing of the two may offer extended tunability in their 2D dispersion map in Fig. \ref{fig7}, especially at the regimes where conventional OEO mode frequencies are hard to generate. 

\begin{figure}[htb]
 \centering
 \includegraphics[width=3.4 in]{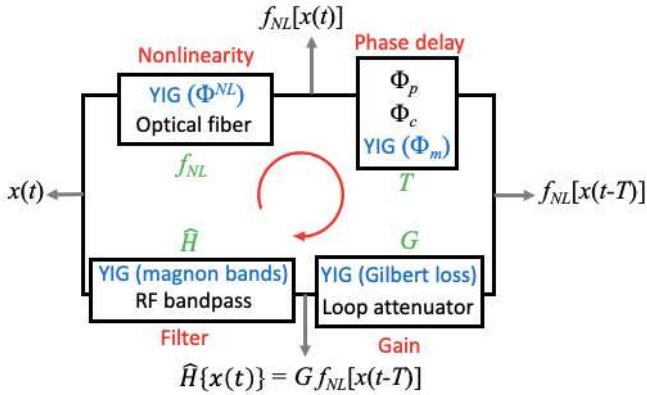}
 \caption{Time-domain block diagram a typical MOEO. The variable \textit{x}(\textit{t}) circulates in the clockwise direction and is subjected to the four characteristic elements of the loop, i.e. nonlinearity, time-delay, gain, and linear filter. Notably, the YIG magnonic component plays a role in all four elements due to its respective properties introduced above. }
 \label{fig10}
\end{figure}

In summary, we report the construction and characterization of a comprehensive MOEO system based on 1550-nm photonics and YIG magnonics. The architecture of the feedback loop, following the time-domain block diagram illustrated in Fig. \ref{fig10}, is characterized by the four essential elements, namely, the nonlinearity, time-delay, gain, and spectral filtering. The dynamics of the system can be tracked by a dynamical variable \textit{x}(\textit{t}). In particular, as compared to conventional OEO systems, the YIG component effectively introduces contributions to all four key elements of the feedback loop, through the YIG's respective magnonic characteristics. Figure 10 indicates that the hybrid optoelectronic systems may be another platform for exploring the advantages of YIG magnonics. 

The MOEO system exhibits a rich parameter space involving synergistic control from the photonic, electronic, and magnonic parts. Taking advantage of the spin wave dispersion of YIG, the frequency self-generation as well as the related nonlinear processes become sensitive to the external magnetic field. Besides known as a narrowband filter, the YIG spin wave modes can be controlled to mix with the OEO modes to generate harmonic beating modes. With the ultrahigh sensitivity \textcolor{black}{of the oscillator frequencies and harmonics} and versatile tunability \textcolor{black}{under external perturbations such as heat, strain, electric field, and spin-torques}, the MOEO system may find usefulness in sensing applications in magnetism and spintronics beyond optoelectronics and photonics.

\textcolor{black}{ \section*{Acknowledgment}
This work, including apparatus buildup, experimental measurements, and data analysis, was supported by U.S. National Science Foundation under Grants No. ECCS-1933301 and ECCS-1941426. Work at Argonne, including sample holder fabrication and spin wave modeling, was supported by U.S. DOE, Office of Science, Materials Sciences and Engineering Division. We are extremely grateful for Andrii Chumak for providing the testing YIG delayline and Vasyl Tyberkevych, Andrei Slavin, Aaron Hagerstrom, and Mingzhong Wu for helpful discussions.  }

\textcolor{black}{ \section*{Data availability}
The data that support the findings of this study are available from the corresponding author upon reasonable request.}\\

%\clearpage


%merlin.mbs apsrev4-1.bst 2010-07-25 4.21a (PWD, AO, DPC) hacked
%Control: key (0)
%Control: author (72) initials jnrlst
%Control: editor formatted (1) identically to author
%Control: production of article title (-1) disabled
%Control: page (0) single
%Control: year (1) truncated
%Control: production of eprint (0) enabled
\begin{thebibliography}{0}%
\makeatletter
\providecommand \@ifxundefined [1]{%
 \@ifx{#1\undefined}
}%
\providecommand \@ifnum [1]{%
 \ifnum #1\expandafter \@firstoftwo
 \else \expandafter \@secondoftwo
 \fi
}%
\providecommand \@ifx [1]{%
 \ifx #1\expandafter \@firstoftwo
 \else \expandafter \@secondoftwo
 \fi
}%
\providecommand \natexlab [1]{#1}%
\providecommand \enquote  [1]{``#1''}%
\providecommand \bibnamefont  [1]{#1}%
\providecommand \bibfnamefont [1]{#1}%
\providecommand \citenamefont [1]{#1}%
\providecommand \href@noop [0]{\@secondoftwo}%
\providecommand \href [0]{\begingroup \@sanitize@url \@href}%
\providecommand \@href[1]{\@@startlink{#1}\@@href}%
\providecommand \@@href[1]{\endgroup#1\@@endlink}%
\providecommand \@sanitize@url [0]{\catcode `\\12\catcode `\$12\catcode
  `\&12\catcode `\#12\catcode `\^12\catcode `\_12\catcode `\%12\relax}%
\providecommand \@@startlink[1]{}%
\providecommand \@@endlink[0]{}%
\providecommand \url  [0]{\begingroup\@sanitize@url \@url }%
\providecommand \@url [1]{\endgroup\@href {#1}{\urlprefix }}%
\providecommand \urlprefix  [0]{URL }%
\providecommand \Eprint [0]{\href }%
\providecommand \doibase [0]{http://dx.doi.org/}%
\providecommand \selectlanguage [0]{\@gobble}%
\providecommand \bibinfo  [0]{\@secondoftwo}%
\providecommand \bibfield  [0]{\@secondoftwo}%
\providecommand \translation [1]{[#1]}%
\providecommand \BibitemOpen [0]{}%
\providecommand \bibitemStop [0]{}%
\providecommand \bibitemNoStop [0]{.\EOS\space}%
\providecommand \EOS [0]{\spacefactor3000\relax}%
\providecommand \BibitemShut  [1]{\csname bibitem#1\endcsname}%
\let\auto@bib@innerbib\@empty
%</preamble>
\end{thebibliography}%


\begin{thebibliography}{19}

\bibitem{markus_rmp2014} M. Aspelmeyer, T. J. Kippenberg, and F. Marquardt, "Cavity Optomechanics", Rev. Mod. Phys. \textbf{86}, 1391 (2014). 

\bibitem{mikhail_rmp2006} M. I. Rabinovich, P. Varona, A. I. Selverston, and H. D. I. Abarbanel, "Dynamical principles in neuroscience", Rev. Mod. Phys. \textbf{78}, 1213 (2006). 

\bibitem{hassan_rmp2017} H. Aref, \textit{et al.}, "Frontiers of chaotic advection", Rev. Mod. Phys. \textbf{89}, 025007 (2017). 

\bibitem{argyris_nature2005} A. Argyris, D. Syvridis, L. Larger, V. Annovazzi-Lodi, P. Colet, I. Fischer, J. Garcia-Ojalvo, C. R. Mirasso, L. Pesquera, and K. A. Shore, "Chaos-based communications at high bit rates using commercial fibre-optics links", Nature. \textbf{438}, 343 (2005).

\bibitem{varela_natrevneuro2001} F. Varela, J. Lachaux, E. Rodriguez, and J. Martinerie, "The brainweb: Phase synchronizatin and large-scale integration", Nat. Rev. Neurosci. \textbf{2}, 229 (2001).

\bibitem{haken_ZPhys1963} H. Haken and H. Sauermann, "Frequency shifts of laser modes in solid state and gaseous systems", Z. Phys. \textbf{176}, 47 (1963).

\bibitem{grollier_jap2018} J. Grollier, S. Guha, H. Ohno, and I. K. Schuller, "Preface to Special Topic: New Physics and Materials for Neuromorphic Computation", J. Appl. Phys. \textbf{124}, 151801 (2018), and the topical articles therein. 

\bibitem{chernilov_pre1995} A. Chernikov and G. Schmidt, "Conditions for synchronization in Josephson-junction arrays", Phys. Rev. E \textbf{52}, 3415 (1995). 

\bibitem{slavin_ieee2009} A. Slavin and V. S. Tiberkevich, "Nonlinear auto-oscillator theory of microwave generation by spin-polarized current", IEEE Transaction on Magnetics \textbf{45},
1875 (2009).

\bibitem{ruotolo_nnano2009} A. Ruotolo, \textit{et al}. "Phase-locking of magnetic vortices mediated by antivortices", Nature Nanotechnology \textbf{4},528 (2009).

\bibitem{grollier_prb2006} J. Grollier, V. Cros, and A. Fert, "Synchronization of spin-transfer oscillators driven by stimulated microwave currents", Phys. Rev. B \textbf{73} 060409(R) (2006).

\bibitem{vasyl_apl2009} V. S. Tiberkevich, A. Slavin, E. Bankowski, and G. Gerhart, "Phase-locking and frustration in an array of nonlinear spin-torque nano-oscillators", Appl. Phys. Lett.
\textbf{95}, 262505 (2009).

\bibitem{yao_ieee1996} X. S. Yao and L. Maleki, "Optoelectronic Oscillator for Photonic Systems", IEEE Journal of Quantum Electronics \textbf{32}, 1141 (1996). 

\bibitem{miguel_rmp2013} M. C. Soriano, J. Garcia-Ojalvo, C. R. Mirasso, and I. Fischer, "Complex photonics: Dynamics and applications of delay-coupled semiconductors lasers", Rev. Mod. Phys. \textbf{85}, 421 (2013). 

\bibitem{larger_rtsa2013} L. Larger, "Complexity in electro-optic delay dynamics: modelling, design and applications", Phil Trans R Soc A \textbf{371}, 20120464 (2013). 

\bibitem{wang_prl2011} Z. Wang, A. Hagerstrom, J. Q. Anderson, W. Tong, M. Wu, L. D. Carr, R. Eykholt, and B. Kalinikos, "Chaotic spin-wave solitons in magnetic film feedback rings", Phys. Rev. Lett. \textbf{107}, 114102 (2011).

\bibitem{wu_ssp2010} M. Wu, "Nonlinear spin waves in magnetic film feedback rings", Solid State Physics \textbf{62}, 163 (2010). 

\bibitem{lute_npho} L. Maleki, "The optoelectronic oscillator", Nature Photonics \textbf{5}, 728 (2011). 

\bibitem{yanne_rmp2019} Y. K. Chembo, D. Brunner, M. Jacquot, and L. Larger, "Optoelectronic oscillators with time-delayed feedback", Rev. Mod. Phys.  \textbf{91}, 035006 (2019). 

\bibitem{harder_ssp2018} M. Harder and C. -M. Hu, "Cavity Spintronics: An Early Review of Recent Progress in the Study of Magnon-Photon Level Repulsion", Solid State Physics, \textbf{69}, 47 (2018). R. Stamps and R. Camley (Ed.), Academic Press.

\bibitem{huebl_prl2013} H. Huebl, C. W. Zollitsch, J. Lotze, F. Hocke, M. Greifenstein, A. Marx, R. Gross, and S. T. B. Goennenwein, "High Cooperativity in Coupled Microwave Resonator Ferrimagnetic Insulator Hybrids", Phys. Rev. Lett. \textbf{111}, 127003 (2013).

\bibitem{xufeng_prl2014} X. Zhang, C. -L. Zou, L. Jiang, and H. X. Tang, "Strongly Coupled Magnons and Cavity Microwave Photons", Phys. Rev. Lett. \textbf{113}, 156401 (2014).

\bibitem{yili_prl2019}Y. Li, \textit{et al.}, "Strong coupling between magnons and microwave photons in resonant ferromagnet-superconductor thin-film devices", Phys. Rev. Lett. \textbf{123}, 107701 (2019). 

\bibitem{luqiao_prl2019} J. T. Hou and L. Liu, "Strong coupling between microwave photons and nanomagnet magnons", Phys. Rev. Lett. \textbf{123}, 107702 (2019). 

\textcolor{black}{\bibitem{yili_jap2020} Y. Li, W. Zhang, V. Tyberkevych, W. -K. Kwok, A. Hoffmann, V. Novosad, "Hybrid magnonics: Physics, circuits, and applications for coherent information processing", J. Appl. Phys. \textbf{128}, 130902 (2020).} 

\bibitem{chumak_nphys2015} A. V. Chumak, V. I. Vasyuchka, A. A. Serga, and B. Hillebrands, "Magnon spintronics", Nature Phys. \textbf{11}, 453 (2015). 

\bibitem{chumak_jphysd2017} A. V. Chumak, A. A. Serga, and B. Hillebrands, "Magnonic crystals for data processing", J. Phys. D: Appl. Phys. \textbf{50}, 244001 (2017). 

\bibitem{grundler_jphysd2010} V. V. Kruglyak, S. O. Demokritov, and D. Grundler, "Magnonics", J. Phys. D: Appl. Phys. \textbf{43}, 260301 (2010). 

\bibitem{slavin_ncomm2010} A. V. Chumak, V. S. Tiberkevich, A. D. Karenowska, A. A. Serga, J. F. Gregg, A. N. Slavin, and B. Hillebrands, "All-linear time reversal by a dynamic artificial crystal", Nature Commun. \textbf{1}, 141 (2010). 

\bibitem{platow_prb1998} W. Platow, A. N. Anisimov, G. L. Dunifer, M. Farle, and K. Baberschke, "Correlations between ferromagnetic-resonance linewidths and sample quality in the study of metallic ultrathin films", Phys. Rev. B \textbf{58}, 5611 (1998).

\bibitem{heinrich_prb2004} G. Woltersdorf and B. Heinrich, "Two-magnon scattering in a self-assembled nanoscale network of misfit dislocations", Phys. Rev. B \textbf{69}, 184417 (2004).

\bibitem{aaron_prl2009} A. M. Hagerstrom, W. Tong, M. Wu, B. A. Kalinikos, and R. Eykholt, "Excitation of Chaotic Spin Waves in Magnetic Film Feedback Rings through Three-Wave Nonlinear Interactions", Phys. Rev. Lett. \textbf{102}, 207202 (2009).

\bibitem{lopes_pre1996} S. R. Lopes and A. C. -L. Chian, "Controlling chaos in nonlinear three-wave coupling", Phys. Rev. Lett. \textbf{54}, 170 (1996).

\bibitem{marsh_prb2012} J. Marsh and R. E. Camley, "Two-wave mixing in nonlinear magnetization dynamics: A perturbation expansion of the Landau-Lifshitz-Gilbert equation", Phys. Rev. B \textbf{86}, 224405 (2012). 

\bibitem{camley_prb2014} R. E. Camley, "Three-magnon processes in magnetic nanoelements: Quantization and localized mode effects", Phys. Rev. B \textbf{89}, 214402 (2014). 

\bibitem{bloom_apl1977} D. M. Bloom and G. C. Bjorklund, "Conjugate wave-front generation and image reconstruction by four-wave mixing", Appl. Phys. Lett. \textbf{31}, 592 (1977). 

\bibitem{serga_prl2005} A. A. Serga, B. Hillebrands, S. O. Demokritov, A. N. Slavin, P. Wierzbicki, V. Vasyuchka, O. Dzyapko, and A. Chumak, "Parametric Generation of Forward and Phase-Conjugated Spin-Wave Bullets in Magnetic Films", Phys. Rev. Lett. \textbf{94}, 167202 (2005). 

\bibitem{celinski_apl2011} Y. Khivintsev, J. Marsh, V. Zagorodnii, I. Harward, J. Lovejoy, P. Krivosik, R. E. Camley, and Z. Celinski, "Nonlinear amplification and mixing of spin waves in a microstrip geometry with metallic ferromagnets", Appl. Phys. Lett. \textbf{98}, 042505 (2011). 

\bibitem{khitun_apl2008} M. Bao, A. Khitun, Y. Wu, J.-Y. Lee, K. L. Wang, and A. P. Jacob, "Coplanar waveguide radio frequency ferromagnetic parametric amplifier", Appl. Phys. Lett. \textbf{93}, 072509 (2008). 

\bibitem{vitko_jpcs2018} V.V. Vitko, A. A. Nikitin, A. B. Ustinov, and B. A. Kalinikos, "A Theoretical Model of Dual Tunable Optoelectronic Oscillator", Journal of Physics: Conf. Series \textbf{1038}, 012106 (2018). 

\bibitem{nikitin_emc2017} A.A. Nikitin, V.V. Vitko, A.V. Kondrashov, A.B. Ustinov, and B.A. Kalinikos, "Theory of Resonant Frequency Spectrum of Tunable Multi-loop Spin-wave Optoelectronic Oscillators", Proceedings of the 47th European Microwave Conference p1108-11 (2017). 

\bibitem{ustinov_metals2018} A. B. Ustinov, A. V. Kondrashov, A. A. Nikitin, A. V. Drozdovskii, and B. A. Kalinikos, "Self-Generation of Chaotic Microwave Signal in Spin Wave Optoelectronic Generator", Physics of the Solid State, \textbf{60}, 2127 (2018).

\bibitem{ustinov_magnlett2015} A. B. Ustinov, A. A. Nikitin, and Boris A. Kalinikos, "Magnetically Tunable Microwave Spin-Wave Photonic Oscillator", IEEE Magn. Lett. \textbf{6}, 3500704 (2015).

\bibitem{vitko_magnlett2018} V. V. Vitko, A. A. Nikitin, A. B. Ustinov,  and Boris A. Kalinikos, "Microwave Bistability in Active Ring Resonators With Dual Spin-Wave and Optical Nonlinearities", IEEE Magn. Lett. \textbf{9}, 3506304 (2018).

\bibitem{thatec} https://www.thatec-innovation.com/

\bibitem{lim2019direct} J. Lim, W. Bang, J. Trossman, A. Kreisel, M. B. Jungfleisch, A. Hoffmann, C. C. Tsai, and J. B. Ketterson, ``Direct detection of multiple backward volume modes in yttrium iron garnet at micron scale wavelengths'', Phys. Rev. B \textbf{99}, 014435 (2019).

\bibitem{kreisel2009microscopic} A. Kriesel, F. Sauli, L. Bartosch, and P. Kopietz,  ``Microscopic spin-wave theory for yttrium-iron garnet films'', The European Physical Journal B \textbf{71}, 59 (2009).

\bibitem{watt_prap2020} S. Watt and M. Kostylev, "Reservoir cmoputing using a spin-wave delay-line active-ring resonator based on Yttrium-Iron-Garnet film", Phys. Rev. Appl. \textbf{13}, 034057 (2020). 

\bibitem{damon1961magnetostatic} R. W. Damon and J. R. Eshbach,  ``Magnetostatic modes of a ferromagnet slab'', Journal of Physics and Chemistry of Solids \textbf{19}, 308-320 (1961).

\bibitem{lim2019magnetostatic}  J. Lim, W. Bang, J. Trossman, D. Amanov, C. C. Tsai, A. Hoffmann, and J. B. Ketterson, ``Magnetostatic spin-waves in an yttrium iron garnet thin film: Comparison between theory and experiment for arbitrary field directions'',Jour. of App. Phys. \textbf{126}, 243906 (2019).

\bibitem{lim2018forward} J. Lim, W. Bang, J. Trossmann, D. Amanov, and J. B. Ketterson, ``Forward volume and surface magnetostatic modes in an yttrium iron garnet film for out-of-plane magnetic fields: Theory and experiment'',AIP Adv. \textbf{8}, 056018 (2018).

\bibitem{vasyl_srep2014} V. S. Tiberkevich, R. S. Khymyn, H. X. Tang, and A. Slavin, "Sensitivity to external signals and synchronization properties of a non-isochronous auto-oscillator with delayed feedback", Sci. Rep. \textbf{4}, 3873 (2014). 


\bibitem{xiong_srep2020} Y. Xiong, \textit{et al.} "Probing magnon–magnon coupling in exchange coupled Y3Fe5O12/Permalloy bilayers with magneto-optical effects", NPG Sci. Rep. \textbf{10}, 12548 (2020). 

\textcolor{black}{\bibitem{yili_YIGPy} Y. Li, \textit{et al.} "Coherent Spin Pumping in a Strongly Coupled Magnon-Magnon Hybrid System", Phys. Rev. Lett. \textbf{124}, 117202 (2020). } 


\bibitem{riou_prap2019} M. Riou, J. Torrejon, B. Garitaine, F. Abreu Araujo, P. Bortolotti, V. Cros, S. Tsunegi, K. Yakushiji, A. Fukushima, H. Kubota, S. Yuasa, D. Querlioz, M.D. Stiles, and J. Grollier, "Temporal Pattern Recognition with Delayed-Feedback Spin-Torque Nano-Oscillators", Phys. Rev. Applied \textbf{12}, 024049 (2019). 

\bibitem{kalinikos_jpc1986} B. A. Kalinikos and A. N. Slavin, "Theory of dipole-exchange spin wave spectrum for ferromagnetic films with mixed exchange boundary conditions", J. Phys. C: Solid State Physics \textbf{19}, 7013 (1986).

\bibitem{demokritov_physrep2001} S. O. Demokritov, B. Hillebrands, and A. Slavin, "Brillouin light scattering studies of confined spin waves: linear and nonlinear confinement", Physics Reports, \textbf{348}, 441 (2001). 

\bibitem{gui_prl2007} Y. S. Gui, N. Mecking, and C. -M. Hu, "Quantized Spin Excitations in a Ferromagnetic Microstrip from Microwave Photovoltage Measurements", Phys. Rev. Lett. \textbf{98}, 217603 (2007). 

%\bibitem{slavin_IEEE2009} Andrei Slavin and Vasil Tiberkevich, "Nonlinear Auto-Oscillator Theory of Microwave Generation by Spin-Polarized Current", IEEE Trans. Magn. \textbf{45}, {1875}, (2009).

\bibitem{muduli_PRB2010} P. K. Muduli, Ye. Pogoryelov, S. Bonetti, G. Consolo, Fred Mancoff,  and J. \AA{}kerman, "Nonlinear frequency and amplitude modulation of a nanocontact-based spin-torque oscillator", Phys. Rev. B \textbf{81}, {140408}, (2010).

\bibitem{ebels_APL2014} M. Quinsat, F. Garcia-Sanchez, A. S. Jenkins,  V. S. Tiberkevich, A. N. Slavin, L. D. Buda-Prejbeanu, A. Zeltser, J. A. Katine, B. Dieny, M.-C. Cyrille, and U. Ebels, "Modulation bandwidth of spin torque oscillators under current modulation", Appl. Phys. Lett. \textbf{105}, {152401}, (2014).

\bibitem{grundler_prap2019} K. An, V.S. Bhat, M. Mruczkiewicz, C. Dubs, and D. Grundler, "Optimization of Spin-Wave Propagation with Enhanced Group Velocities by Exchange-Coupled Ferrimagnet-Ferromagnet Bilayers", Phys. Rev. Applied \textbf{11}, 034065 (2019). 


\bibitem{melkov_jap2005} G. A. Melkov, V. I. Vasyuchka, A. V. Chumak, and A. Slavin, "Double-wave-front reversal of dipole-exchange spin waves in yttrium-iron garnet films", J. Appl. Phys. \textbf{98}, 074908 (2005). 

\bibitem{melkov_prb2009} G. A. Melkov, Yu. V. Koblyanskiy, R. A. Slipets, A. V. Talalaevskij, and A. N. Slavin, "Nonlinear interactions of spin waves with parametric pumping in permalloy metal films", Phys. Rev. B \textbf{79} 134411 (2009). 

\bibitem{inglis_jmmm2019} A. Inglis and J. F. Gregg, "Onset of spin wave time-domain fractals in a dynamic artificial crystal", J. Magn. Magn. Mater. \textbf{495} 165868 (2020). 



\end{thebibliography}
\end{document}